\documentclass[11pt]{article}

\usepackage[letterpaper,margin=1in]{geometry}
\usepackage[T1]{fontenc}
\usepackage{lmodern}
\usepackage{amsmath,amssymb,amsthm}
\usepackage{graphicx}
\usepackage{booktabs}
\usepackage{microtype}
\usepackage{xcolor}
\usepackage{xurl}
\usepackage[numbers,sort&compress]{natbib}

\definecolor{qnavy}{HTML}{16324F}
\definecolor{qteal}{HTML}{2A6F77}

\usepackage[colorlinks=true,linkcolor=qnavy,citecolor=qteal,urlcolor=qteal]{hyperref}
\hypersetup{
  pdftitle={Direct quantum simulation of compressible Navier--Stokes flow via finite homotopy linearisation},
  pdfauthor={Ming-Yang Tan, Guo-Ping Guo, Zhao-Yun Chen}
}

\providecommand{\doi}[1]{\href{https://doi.org/#1}{\nolinkurl{doi:#1}}}

\makeatletter
\def\ps@headings{%
  \let\@oddfoot\@empty
  \def\@oddhead{{\small\color{qnavy!75!black}\rightmark}\hfil{\small\thepage}}%
  \let\@mkboth\markboth
  \def\sectionmark##1{%
    \markright{%
      \ifnum \c@secnumdepth >\m@ne
        \thesection\quad
      \fi
      ##1}}}
\makeatother

\numberwithin{equation}{section}

\theoremstyle{plain}
\newtheorem{theorem}{Theorem}

\newtheorem{proposition}{Proposition}

\title{\textbf{Direct quantum simulation of compressible Navier--Stokes flow via finite homotopy linearisation}}
\author{%
  Ming-Yang Tan$^{1}$,\quad Guo-Ping Guo$^{1,2,3,4}$,\quad and\quad Zhao-Yun Chen$^{3,}$\thanks{Corresponding author: Zhao-Yun Chen, \href{mailto:chenzhaoyun@iai.ustc.edu.cn}{\texttt{chenzhaoyun@iai.ustc.edu.cn}}.}\\[0.55em]
  \parbox{0.96\textwidth}{\centering\footnotesize
    $^{1}$ Laboratory of Quantum Information, University of Science and Technology of China,\\
    Hefei, Anhui 230026, China\\
    $^{2}$ Anhui Province Key Laboratory of Quantum Network, University of Science and Technology of China,\\
    Hefei, Anhui 230026, China\\
    $^{3}$ Institute of Artificial Intelligence, Hefei Comprehensive National Science Center, Hefei, Anhui 230088, China\\
    $^{4}$ Origin Quantum Computing Technology (Hefei) Co., Ltd., Hefei, Anhui 230026, China}%
}
\date{}

\begin{document}
\maketitle
\thispagestyle{plain}

\begin{abstract}
We develop a direct quantum simulation algorithm for finite homotopy approximations to the non-isothermal compressible Navier--Stokes equations. A primitive-variable formulation gives polynomial residuals in the fields and their spatial and time derivatives, including density-weighted time derivatives. On a periodic grid, applying each spatial stencil uniformly to every factor in a shifted product preserves the dynamics of the finite homotopy hierarchy in an affine linearisation. Because each deformation equation depends only on lower homotopy orders, the lifted generator matrix is strictly block lower-triangular and hence nilpotent. The resulting finite-polynomial solution evaluates the state at the final time directly without time-stepping discretisation error, while explicit short-time convergence conditions bound the homotopy truncation error on a fixed grid. In two-dimensional Taylor--Green flow, initial pressure balance removes the leading longitudinal acceleration and delays the onset of dilatational kinetic energy from quadratic to quartic in time. At Mach numbers $0.2$ and $0.4$, the fifth-order hierarchy reproduces viscous dissipation over a longer interval than pressure--dilatation at the same relative tolerance. We identify auxiliary variables that represent the same evolving product, compile the finite-polynomial evolution into concrete logical quantum circuits, and recover a signed velocity Fourier mode by Hadamard interference. Together, the Taylor--Green energy budgets quantify the time intervals over which a given homotopy order stays within relative accuracy tolerances, while coarse-grid logical circuits verify the signed readout and measure the gate counts of concrete circuit implementations. For higher-order evolution on finer grids, merging coincident stencil terms and combining block-encoded products along paths of strictly increasing homotopy order substantially reduce the predicted normalisation cost.
\end{abstract}

\section{\label{sec:intro}Introduction}

In a compressible vortex, pressure--dilatation exchanges kinetic and internal energy while viscous stress dissipates kinetic energy. The strength and timescale of the first process can change with the initial pressure, even at a fixed initial velocity and identical transport coefficients. Studies of the compressible Taylor--Green problem have documented the growth of compressive motion and its coupling to pressure work \citep{zhang2025taylor}. Here we ask how initial pressure balance and Mach number affect both the onset of compressive motion and the time over which a low-order finite homotopy hierarchy resolves its energy transfer.

The homotopy analysis method (HAM) \citep{liao2003beyond,liao2004homotopy} organises nonlinear evolution into a hierarchy of deformation equations ordered by an embedding parameter. Quantum homotopy formulations have been developed for dissipative quadratic ordinary differential equations \citep{xue2021qhpm} and proposed more generally as the HAM--Schr\"odingerisation framework for  nonlinear partial differential equations \citep{liao2024general}. \citet{xue2025qham} developed product lifting for the Burgers and KdV hierarchies and pointed out cubic terms containing a time derivative in their Navier--Stokes outlook. For non-isothermal compressible flow, the density multiplies the time derivatives of velocity and internal energy, and retaining this mass matrix in the primitive-variable residuals keeps the equations polynomial. We close the resulting hierarchy under  spatial and temporal derivatives by introducing auxiliary product variables, and use its strict lower-order dependence to evaluate the final state directly without introducing a temporal grid.

We derive the fixed-grid product lifting and explicit short-time convergence conditions, and use one-dimensional calculations to check the assembled operator and measure the truncation error. The two-dimensional Taylor--Green calculations compare flows with identical initial velocity fields under two pressure initialisations and resolve their kinetic-energy budgets. The quantum algorithm evaluates the finite-polynomial final-time solution and uses Hadamard interference to recover a signed velocity Fourier mode. The logical circuit implementation measures the resources of concrete propagation and readout circuits, while the Taylor--Green results identify the Mach numbers and initial-pressure profiles over which the chosen homotopy order resolves the energy transfer.

\section{\label{sec:model}The Navier--Stokes equations in polynomial form}

We work in $d\ge1$ space dimensions with periodic boundary conditions. Let $\rho$ be the density, $\boldsymbol{u}=(u_{1},\dots,u_{d})$ the velocity vector, and $e$ the specific internal energy. The pressure, the Newtonian stress tensor, and the Fourier heat-flux vector close the system,
\begin{equation}
p=(\gamma-1)\rho e,\qquad
\boldsymbol{\tau}=\mu\bigl(\nabla \boldsymbol{u}+(\nabla \boldsymbol{u})^{\mathsf T}\bigr)+\lambda\,(\nabla\!\cdot\!\boldsymbol{u})\,\boldsymbol{I}_{d},
\qquad
\boldsymbol{q}=-\kappa\,\nabla e,
\label{eq:closure}
\end{equation}
with ratio of specific heats $\gamma$, dynamic viscosity $\mu$, second viscosity $\lambda$,  coefficient $\kappa$ relating heat flux to the internal-energy gradient, and $d\times d$ identity tensor $\boldsymbol{I}_{d}$.

In the conserved variables $(\rho,\rho\boldsymbol{u},\rho E)$, $E=e+|\boldsymbol{u}|^{2}/2$, the standard conservation statement has fluxes that contain negative powers of the density once the closure~\eqref{eq:closure} is substituted, since $\boldsymbol{u}=(\rho\boldsymbol{u})/\rho$; the residuals are then rational rather than polynomial in the unknowns, and polynomial PDE machinery does not apply directly. The following primitive-variable form restores a polynomial structure in every dimension.

\begin{proposition}[Polynomial primitive residuals]
\label{prop:companion}
The non-isothermal compressible Navier--Stokes equations have the mass, momentum and internal-energy residuals
\begin{subequations}
\label{eq:system}
\begin{align}
N_{\rho}&:\quad \partial_{t}\rho+\boldsymbol{u}\cdot\nabla\rho+\rho\,\nabla\!\cdot\!\boldsymbol{u}=0,
\label{eq:rho}\\
N_{u_{j}}&:\quad \rho\,\partial_{t}u_{j}+\rho\,\boldsymbol{u}\cdot\nabla u_{j}+\partial_{x_{j}}p-\bigl(\nabla\!\cdot\!\boldsymbol{\tau}\bigr)_{j}=0,
\qquad j=1,\dots,d,
\label{eq:u}\\
N_{e}&:\quad \rho\,\partial_{t}e+\rho\,\boldsymbol{u}\cdot\nabla e+p\,\nabla\!\cdot\!\boldsymbol{u}-\boldsymbol{\tau}:\nabla\boldsymbol{u}+\nabla\!\cdot\!\boldsymbol{q}=0.
\label{eq:e}
\end{align}
\end{subequations}
Substitution of Eq.~\eqref{eq:closure} expresses each residual as a polynomial of degree at most three in the physical fields and their derivatives. Each factor carries at most one time derivative and at most two spatial derivatives.
\end{proposition}

\begin{proof}
The pressure is quadratic, and the stress and heat flux are linear in derivatives of the primitive variables. The density-weighted material derivatives and the pressure--dilatation term have at most three factors. The stated derivative orders follow directly from the stress divergence and heat-flux divergence.
\end{proof}

To test closure for cubic products with a time derivative of the type discussed by \citet{xue2025qham}, we also use an auxiliary kinetic-energy equation with residual
\begin{equation}\label{eq:k}
 N_k=\sum_{j=1}^d u_j N_{u_j}+\tfrac12|\boldsymbol{u}|^2 N_\rho,\qquad
 k=\tfrac12\rho|\boldsymbol{u}|^2.
\end{equation}
Combining the time terms gives $\partial_t k$. The same combination of the convective terms gives $\nabla\cdot(k\boldsymbol{u})$, so this equation is a differential consequence of the physical mass and momentum equations. Its polynomial expansion contains $(\rho u_j)\partial_t u_j$ and terms of degree four, including $\tfrac12 |\boldsymbol{u}|^2\,\boldsymbol{u}\cdot\nabla\rho$. In the deformation hierarchy the cubic term generates the time-differentiated product of homotopy orders $\{0,1,1\}$ tested in Table~\ref{tab:numeric}. The convergence, flow and quantum-resource benchmarks use only the physical mass, momentum and internal-energy equations, while the auxiliary kinetic-energy equation is reserved for the closure tests.

The one-dimensional reduction, with the auxiliary kinetic-energy equation on the last line, reads

\begin{subequations}
\label{eq:oned}
\begin{align}
&\partial_{t}\rho+u\,\partial_{x}\rho+\rho\,\partial_{x}u=0,\\
&\rho\,\partial_{t}u+\rho u\,\partial_{x}u+\partial_{x}p-\partial_{x}\tau=0,\\
&\rho\,\partial_{t}e+\rho u\,\partial_{x}e+p\,\partial_{x}u-\tau\,\partial_{x}u+\partial_{x}q_{x}=0,\\
&\tfrac{u^{2}}{2}\,\partial_{t}\rho+(\rho u)\,\partial_{t}u+\tfrac{u^{3}}{2}\,\partial_{x}\rho+\tfrac{3}{2}\rho u^{2}\,\partial_{x}u+u\,\partial_{x}p-u\,\partial_{x}\tau=0,
\end{align}
\end{subequations}

with $q_{x}=-\kappa\,\partial_{x}e$ and $\tau=(2\mu+\lambda)\,\partial_{x}u$; the first three equations are the classical one-dimensional system in advective form, and the fourth is Eq.~\eqref{eq:k} expanded.

\section{\label{sec:compare}Related quantum approaches}

Quantum algorithms for nonlinear fluids differ in their time-marching strategies and input--output models. Amplitude-estimation integration produces a classical trajectory~\citep{gaitan2020finding}, whereas steady-state quantum finite volumes combine quantum linear solvers with $\ell^\infty$ tomography and classical updates~\citep{chen2022qfvm}. Subsequent hybrid approaches pair iterative quantum linear solvers with classical computational fluid dynamics outer loops or subspace updates~\citep{chen2024highprecisionqcfd,ye2024hybridcfd}, or advance unsteady compressible flow via implicit finite-volume steps and sparse spectral input--output~\citep{zhuang2025nspathway}. Lattice Boltzmann and variational routes instead rely on intermediate tomography and re-encoding or classical parameter optimisation during evolution~\citep{budinski2022quantum,lubasch2020variational}, including space-time variational formulations tested on nonlinear fluid dynamics~\citep{ma2026vqnls}. For direct linearisation, Carleman methods truncate an infinite tensor embedding; for dissipative quadratic ordinary differential equations, the corrected quantum convergence guarantee assumes both $R<1$ and  the  forcing condition $\|F_0\|\le\|F_2\|$, with complexity also depending on the norm ratio $\|u_{\rm in}\|/\|u(T)\|$~\citep{liu2021efficient}. The accuracy of a finite HAM lift instead depends on convergence of its underlying deformation series, for which Sec.~\ref{sec:defo} gives explicit finite-time sufficient conditions.

Quantum representations of vortex filaments encode coherent structures from a given velocity field for reconstruction and analysis~\citep{zhu2025vortices}. For time evolution, both the encoding of the lifted input vector and the chosen physical observable  determine the cost of propagation and readout.

Among quantum homotopy methods, the closest approach is the frozen--Oseen quantum homotopy analysis method  (FOQHAM)~\citep{yuan2026foqham}, which freezes the semidiscrete Fr\'echet derivative $G$ at a prescribed profile, uses $\partial_t-G$ as the auxiliary linear operator while retaining the known source terms and the nonlinear remainder, closes the hierarchy by product lifting, and reports classical finite-node tests of linear combinations of Hamiltonian simulation (LCHS) for  one- and two-dimensional isothermal compressible flows with equation of state $p=\rho$ and no independent internal-energy evolution equation. By contrast, the present work retains the non-isothermal internal-energy equation and the density-weighted time derivatives, and chooses $L=\partial_t$ as the auxiliary operator. This choice makes the entire lifted generator matrix strictly block lower-triangular in homotopy order, supporting the finite-polynomial solution of Sec.~\ref{sec:cost} without time-stepping discretisation. Beyond direct polynomial evaluation of a nilpotent generator, LCHS, Schr\"odingerisation and contour-based matrix decomposition provide complementary routes to general non-unitary linear dynamics~\citep{an2023lchs,an2026lchs,jin2022schrodingerisation,wang2026cbmd}.

\section[Deformation hierarchy and homotopy ordering]{\label{sec:defo}Deformation hierarchy and homotopy ordering}

The linearisation algorithm of Sec.~\ref{sec:lift} applies to any polynomial system equipped with a strict order grading, in which each unknown factor carries a non-negative homotopy order and every residual monomial in the equation governing an order-$i$ unknown has total order at most $i-1$. The homotopy analysis method produces such a grading for  Eqs.~\eqref{eq:system}, which we now record explicitly.

Embed the system in a scalar homotopy embedding parameter $q\in[0,1]$, written in light italic to distinguish it from the heat-flux vector $\boldsymbol{q}$ of Sec.~\ref{sec:model}. Let $U^{(0)}$ be the initial guess, let $L=\partial_{t}$ be the linear auxiliary operator, and consider the zeroth-order deformation equation
\begin{equation}
(1-q)\,L\big(U(q)-U^{(0)}\big)=q\,h\,N\big(U(q)\big),
\qquad
U(q)=\sum_{j\ge 0}U^{(j)}\,q^{j},
\label{eq:op}
\end{equation}
with convergence-control parameter $h$ and truncation order $m$. Throughout, parenthesised superscripts $(j)$ denote homotopy orders, avoiding confusion with field powers or component subscripts. Expanding in $q$ and matching coefficients gives the deformation equations
\begin{equation}
\partial_{t}U_{a}^{(i)}=h\sum_{n=0}^{i-1}[N_{a}]_{n},
\qquad i=1,\dots,m,
\label{eq:defo}
\end{equation}
where $a$ indexes the field components, the approximation retains $U^{(0)},\ldots,U^{(m)}$, and $[N_{a}]_{n}$ is the coefficient of $q^{n}$ in $N_{a}(U(q))$. Because every monomial of Eqs.~\eqref{eq:system} is a product of field factors and their derivatives, and a factor $U^{(j)}$ contributes homotopy order $j$ regardless of its spatial or temporal derivative orders, each monomial of $[N_{a}]_{n}$ has order sum exactly $n$. Equation~\eqref{eq:defo} is therefore strictly lower-triangular in homotopy order: every nonlinear monomial on the right-hand side of the equation for $U_{a}^{(i)}$ has order sum at most $i-1$. Monomials of order sum zero involve only $U^{(0)}$ and its spatial derivatives, known functions determined by the initial guess; they act as inhomogeneous forcing.

Throughout the paper, we take the initial guess $U^{(0)}$ to be a time-independent spatial field, so its time derivative vanishes identically, $\partial_{t}U^{(0)}\equiv0$, and every residual monomial carrying a zeroth-order time-differentiated factor is dropped from $[N_{a}]_{n}$ as the zero polynomial. Three consequences follow. First, the first-order deformation is driven by the residual evaluated on the initial field, $\partial_{t}U_{a}^{(1)}=h\,N_{a}(U^{(0)})$. Second, a first-order time derivative $\partial_{t}U^{(1)}$ is a \emph{time-independent} data function, $h\,[N]_{0}$, whose own time derivative vanishes. Third, the lifting retains only product variables that remain after eliminating these identically zero terms, so monomials vanishing with $\partial_{t}U^{(0)}$ generate no auxiliary variables.

Boundary and initial conditions enter this deformation before any quantum solver is called. The initial guess $U^{(0)}$ must satisfy the prescribed initial and boundary data, and every deformation increment $U_{a}^{(i)}$, \(i\ge1\), must satisfy the corresponding homogeneous conditions, so that every partial sum satisfies the same data. The periodic stencil enforces periodicity and we impose $U_{a}^{(i)}(0)=0$ for each positive homotopy order. For nonperiodic data one may either put all inhomogeneous boundary values into $U^{(0)}$ or append affine ghost-cell constraints; the second option changes the stencil and must be included in the dimension, sparsity and boundary-oracle accounting.

\subsection{Short-time convergence on a fixed grid}

Consider the semidiscrete physical equations (mass, momentum and internal energy) on a finite periodic grid, and denote the time-independent initial vector by $U^{(0)}$. In our semidiscrete scheme, first spatial derivatives are approximated by the second-order centred difference operator $D$, second spatial derivatives in the viscous and thermal terms are formed as the composition $D^2=D\circ D$ with five-point stencil $(f_{p+2}-2f_p+f_{p-2})/(4\Delta x^2)$, and the discrete pressure gradient is expanded by the product rule as $(\gamma-1)(\rho\,De+e\,D\rho)$ so that the semidiscrete residuals remain polynomial in the primitive variables. Write the semidiscrete residual as $N(U)=M(U)\dot U-F(U)$, where $M$ has unit entries for density and local density entries for velocity and internal energy, and $F$ is the vector-valued polynomial obtained by applying these spatial difference operators to the residual terms. We use the maximum norm $\|\cdot\|_\infty$ on the complexified grid vector space and its associated induced matrix norm.

\begin{theorem}[Explicit short-time convergence bound]\label{thm:short}
Let $U^{(0)}$ have strictly positive density, and choose $h<0$ with $a_0=\|I+hM(U^{(0)})\|_\infty<1$. Choose $r>1$ and $\delta>0$ less than the minimum initial density such that $d_0=1-r(a_0+|h|\delta)>0$. Let $F_\delta$ bound $\|F\|_\infty$ and $K_\delta$ bound the Lipschitz constant of $F$ on the complex maximum-norm ball of radius $\delta$ about $U^{(0)}$. Put
\begin{equation}
 B=\frac{r|h|F_\delta}{d_0},\qquad K=\frac{r|h|K_\delta}{d_0}+
       \frac{(r|h|)^2F_\delta}{d_0^2}.
\end{equation}
If $TB\le\delta$ and $TK<1$, the HAM series converges on $[0,T]$ at $q=1$ to the physical semidiscrete solution. Its order-$m$ tail satisfies
\begin{equation}\label{eq:short-tail}
 \sup_{0\le t\le T}\left\|U(t)-U^{(0)}-\sum_{j=1}^m U^{(j)}(t)\right\|_\infty
 \le \frac{\delta\,r^{-m}}{r-1} .
\end{equation}
\end{theorem}

The constants are computable from the spatial stencil norms and polynomial coefficients. The maximum-norm ball includes grid perturbations near the largest discrete derivative scales, where the norm of the composite second-derivative operator $D^2$ grows as $\Delta x^{-2}$. For a Fourier mode of fixed wave number $k$, the centred first derivative has multiplier of magnitude $|\sin(k\Delta x)|/\Delta x\le |k|$. At fixed order, the trigonometric initial data in our tests generate only finitely many harmonics in the deformation coefficients, so the same multiplier bound applies to them on all grids that resolve these harmonics. This explains why the observed same-grid error can vary weakly with resolution while the worst-case convergence time bound shrinks. A grid-uniform truncation theorem would also require uniform control of the physical solution and the homotopy tail. For example, a sufficient choice of $h$ is $-2/\max\{1,\max_p\rho^{(0)}(p)\}<h<0$, where $\rho^{(0)}$ is the density component of $U^{(0)}$. The auxiliary kinetic-energy equation of Sec.~\ref{sec:model} is driven one-way by the physical increments, and we denote its accumulated homotopy variable by $K_{\mathrm{HAM}}$ to distinguish it from physical kinetic energy. Because the common auxiliary operator $L=\partial_t$ carries no density factor, the velocity and auxiliary kinetic-energy equations share the same scalar multiplier $(1-q)$ on their time derivatives and the same factor $qh$ on their residuals in the zeroth-order deformation. Eliminating the residuals gives $\partial_t K_{\mathrm{HAM}}=\boldsymbol{u}\cdot\partial_t\boldsymbol{u}+(|\boldsymbol{u}|^2/2)\partial_t\rho$ along the deformed trajectory, whose difference from the physical kinetic-energy derivative is $(1-\rho)\boldsymbol{u}\cdot\partial_t\boldsymbol{u}$. We use $K_{\mathrm{HAM}}$ only to test closure of the time-differentiated cubic product, and evaluate physical kinetic energy directly from the reconstructed $\rho$ and $\boldsymbol{u}$. Appendix~\ref{app:convergence} gives the proof and a conditional dissipative extension.

\section[Product lifting with spatial and temporal derivatives]{\label{sec:lift}Product lifting with spatial and temporal derivatives}

\subsection{From one residual term to matrix entries}

Continuing with parenthesised superscripts for homotopy order, let $D$ be the centred first-derivative operator (with $D^2=D\circ D$ for second derivatives). In the one-dimensional momentum residual, the $q^2$ coefficient of $\rho u D u$ is

\begin{align}
 &\rho^{(0)}u^{(1)}D u^{(1)}+\rho^{(1)}u^{(0)}D u^{(1)}+\rho^{(1)}u^{(1)}D u^{(0)}\nonumber\\
 &\quad+\rho^{(0)}u^{(0)}D u^{(2)}+\rho^{(0)}u^{(2)}D u^{(0)}+\rho^{(2)}u^{(0)}D u^{(0)}.
\end{align}

It enters $\dot u^{(3)}$ with prefactor $h$ by Eq.~\eqref{eq:defo}. Introduce the auxiliary product variable $W(p)=u^{(1)}(p)D u^{(1)}(p)$ for the first term. If row and column labels of $A_m$ comprise a product variable index and a grid point, the resulting entry is
\begin{equation}\label{eq:convection-entry}
 (A_m)_{(u^{(3)},p),(W,p)}=h\rho^{(0)}(p)
\end{equation}
for this contribution. The fourth term contributes $\pm h\rho^{(0)}(p)u^{(0)}(p)/(2\Delta x)$ to the columns $(u^{(2)},p\pm1)$. These are the same-point and shifted stencil terms encoded in Appendix~\ref{app:circuits}.

The cubic term $(\rho u)\dot u$ in $N_k$ supplies $\rho^{(0)}u^{(1)}\dot u^{(1)}$ at $q^2$. It gives a second same-point entry, coupling the product variable $u^{(1)}\dot u^{(1)}$ to $k^{(3)}$. Here $\dot u^{(1)}=h[N_u]_0$ is a known grid function, so, for example, $\partial_t(u^{(1)}\dot u^{(1)})=(h[N_u]_0)^2$. At higher order the analogous derivatives are evaluated recursively through the corresponding lower-order equations. A second-order time derivative can thus appear even though every input residual contains at most first-order time derivatives.

\subsection{Product variables and closure}

A product factor is a field component equipped with a homotopy order $j$ and spatial and temporal derivative orders (possibly zero). A product variable index identifies the product of its positive-order factors, while zeroth-order factors contribute known coefficients computed from $U^{(0)}$. An unreduced monomial enumeration retains those zeroth-order labels to trace the originating terms. The compact representation indexes only the positive-order product stored in the lifted vector: it absorbs every first-order time derivative and its spatial derivatives into the time-independent coefficients, discards time derivatives above the temporal polynomial degree of an increment, and reclassifies any term left with at most one evolving factor. Terms sharing the same evolving product then map to the same column. All contributions are still summed, and the reduction preserves the strict lower-order dependence of every equation. In the continuum, a product of differentiated factors is also the diagonal restriction of a product on independent copies of space-time, after applying each derivative to its own copy, and evaluating on the diagonal is linear in the function on the product space.

The evolution of a product variable follows from the time product rule. For a factor $\partial_x^\alpha\partial_t^\beta U^{(j)}$, we use $\partial_x^\alpha\partial_t^\beta R^{(j)}$ for its rate, where $\dot U^{(j)}=R^{(j)}$, and expand time derivatives by the product rule. A time derivative on a time-independent factor vanishes. Any generated factor with a higher-order time derivative may be retained or eliminated by substituting $\partial_t^{\beta-1}R^{(j)}$ for $\partial_t^\beta U^{(j)}$ and applying its spatial derivative. Each substitution replaces homotopy order $j$ by a sum of orders at most $j-1$. In particular, induction in $j$ shows that $U^{(j)}(t)$ has polynomial degree in time  at most $j$ on a fixed grid, so time derivatives beyond that degree vanish on consistent states.

In the continuum, spatial derivatives of $R^{(j)}$ can be expanded by the spatial Leibniz rule. On a discrete grid, a centred difference stencil does not satisfy the continuum Leibniz rule; instead, applying the stencil to a product evaluates the entire product at each shifted grid point,
\begin{equation}\label{eq:shifted-product}
 D^\alpha\!\left(\prod_{i=1}^r F_i\right)(p)=
 \sum_\eta d^{(\alpha)}_\eta\prod_{i=1}^r F_i(p+\eta).
\end{equation}
The shift $\eta$ is applied to every factor of the replaced monomial, while factors outside that monomial keep their original offsets. A product variable index on the grid can therefore include relative offsets. This distinction matters for comparing the lifting with the semidiscrete equations, because centred differences do not obey the continuum Leibniz rule.

\begin{theorem}[Termination and exactness]\label{thm:term}
For a finite derivative-bearing polynomial hierarchy with time-independent zeroth-order data and strict lower-order dependence, product lifting terminates at each finite $m$. For smooth continuum states, or on the fixed grid with Eq.~\eqref{eq:shifted-product}, the lifted affine right-hand side agrees component by component with the time derivative of the consistent product values. On the fixed grid the consistent states form an invariant graph over the base deformation variables.
\end{theorem}

\begin{proof}
Assign to each base variable its homotopy order, and to each auxiliary product variable the sum of the homotopy orders of its positive-order factors. Each substitution of a lower-order equation strictly decreases that sum. A term with zero positive-order factors contributes to the inhomogeneous source $b_m$, and a term with one factor becomes a linear coupling to a single unknown after resolving any time derivative through lower-order equations. A term with at least two positive-order factors points to an auxiliary product variable of strictly lower order sum. Derivative expansion and each finite stencil create finitely many terms, so a variable of order sum $j$ has substitution depth at most $j$. Each substitution is an identity on a consistent deformation state, including  Eq.~\eqref{eq:shifted-product}. The exact time derivative of its product values thus solves the affine system. Uniqueness for the finite-dimensional affine equation gives the invariant-graph statement.
\end{proof}

For any fixed $m$, the number of possible stencil offsets and the number of product variable indices per grid point are bounded independently of the resolution, and the finite substitution depth bounds both row and column sparsity. The size of the resulting list of auxiliary variables depends on $m$, on the residual polynomials, and on how higher-order derivatives are resolved, while grid size still affects the numerical coefficients and the cost of preparing their data.

\section{\label{sec:verify}One-dimensional verification and physical truncation error}

The independent reference evaluates each $U^{(j)}$ as a time polynomial by multiplying the lower-order time polynomials in the deformation equations and integrating their coefficients, applying spatial stencils directly to grid data. To verify the lifting, we differentiate its consistent products using these same independent reference polynomials and compare every component with the assembled affine right-hand side $A_m y + b_m$. We also compare the full lifted vector at the final time against the reference products. The finite-grid initialisation includes products of time-differentiated factors, which can be nonzero even when all positive-order base increments vanish at $t=0$.

\begin{table}
\caption{\label{tab:numeric}One-dimensional product lifting at $n=8$ using Eq.~\eqref{eq:shifted-product}. Columns labelled Physical omit the auxiliary kinetic-energy equation; columns labelled Auxiliary include it. Each variable block holds $n$ grid values. Rowwise residual and final-time error are maximum absolute deviations for the augmented system over $t=0,0.004,0.02$.}
\vspace{4pt}
\centering
\small
\begin{tabular}{crrrrrr}
\toprule
 & \multicolumn{2}{c}{Physical} & \multicolumn{2}{c}{Auxiliary} & Rowwise & Final-time\\
\cmidrule(lr){2-3}\cmidrule(lr){4-5}
$m$ & Blocks & Max row nnz & Blocks & Max row nnz & residual & error\\
\midrule
1 & 3 & 0 & 4 & 0 & 0 & 0\\
2 & 6 & 11 & 8 & 11 & $2.78\times 10^{-17}$ & $2.17\times 10^{-19}$\\
3 & 25 & 28 & 41 & 38 & $2.78\times 10^{-17}$ & $8.67\times 10^{-19}$\\
4 & 331 & 52 & 492 & 96 & $8.88\times 10^{-16}$ & $2.78\times 10^{-17}$\\
5 & 4562 & 90 & 5881 & 193 & $1.07\times 10^{-14}$ & $2.78\times 10^{-16}$\\
\bottomrule
\end{tabular}
\end{table}

\begin{figure}
\centering
\includegraphics[width=0.82\textwidth]{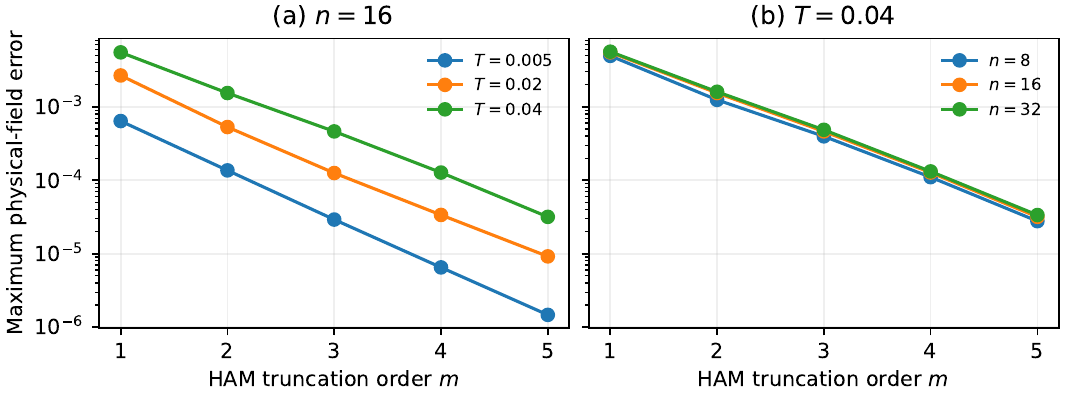}
\caption{\label{fig:ham-error}Maximum physical-field error of the one-dimensional HAM truncation compared with direct reference integration on the same periodic grid. Left, dependence on final time at fixed $n=16$. Right, dependence on the number of cells at fixed $T=0.04$.}
\end{figure}

The physical error experiment integrates the three original semidiscrete equations using an adaptive eighth-order Runge--Kutta integrator (DOP853) after dividing the momentum and internal-energy equations by the density. It uses the same centred spatial scheme as the HAM series (with $D^2=D\circ D$), with $\gamma=1.4$, $\mu=0.01$, $\lambda=0$, $\kappa=0.01$, $h=-0.8$, and initial profiles $\rho_0=1+0.05\sin(2\pi x)$, $u_0=0.1\sin(2\pi x)$ and $e_0=2.5+0.05\cos(2\pi x)$. Figure~\ref{fig:ham-error} measures the maximum difference in $\rho,u,e$ at the final time and isolates the HAM truncation error for the chosen spatial discretisation. For $r=1.5$ and $\delta=0.1$, estimating the polynomial and its Lipschitz constant by sums of stencil norms gives, by Theorem~\ref{thm:short}, an explicit convergence time bound of $T=0.0092$ at $n=16$. The maximum change on refining the reference tolerance was $3.4\times10^{-13}$. At $n=16$ and $T=0.04$, the error decreases from $5.51\times10^{-3}$ at first order  to $3.18\times10^{-5}$ at fifth order.

\subsection{Cost of the auxiliary degrees of freedom}

The unreduced variable list counts monomials with distinct derivative orders, grid offsets and zeroth-order data labels separately. Table~\ref{tab:compact} measures the effect of identifying identical evolving products and absorbing time-independent first-order time derivatives into known coefficients. At fifth order, this reduction lowers the one-dimensional physical system from 4547 to 787 auxiliary product blocks. The physical totals in Table~\ref{tab:numeric} also include $3m$ positive-order base-increment blocks, one per physical field at each order; at $m=5$, the 4547 unreduced product blocks and  15 base-increment blocks give the total of 4562. The maximum number of nonzeros per column  drops by a factor of four, while the maximum per row  changes only slightly, and the normalisation $\alpha$ of the factorised stencil block encoding also decreases. The rowwise algebraic residual for the compact representation  against independently differentiated products stays below $9\times10^{-16}$ at fifth order.

\begin{table}
\caption{\label{tab:compact}Reduction of auxiliary product variables for the physical equations on an eight-point one-dimensional periodic grid at $T=0.02$. The operator is time independent. The displayed $\alpha$ is the sum of individual stencil-term bounds for the generator before merging.}
\vspace{4pt}
\centering
\small
\begin{tabular}{c r r c c r r}
\toprule
 & \multicolumn{2}{c}{Product blocks} & \multicolumn{2}{c}{Max nonzeros (row/col)} & \multicolumn{2}{c}{Unmerged stencil bound $\alpha$}\\
\cmidrule(lr){2-3}\cmidrule(lr){4-5}\cmidrule(lr){6-7}
$m$ & Unreduced & Compact & Unreduced & Compact & Unreduced & Compact\\
\midrule
3 & 16 & 10 & 28/39 & 27/34 & 251.1 & 226.2\\
4 & 319 & 111 & 52/325 & 50/171 & 7138.6 & 5696.3\\
5 & 4547 & 787 & 90/3746 & 87/934 & 159188.2 & 89006.4\\
\bottomrule
\end{tabular}
\end{table}

For context, a full symmetric Carleman basis of polynomial degree at most $r$ on $N_0$ grid unknowns has $\sum_{j=1}^{r}\binom{N_0+j-1}{j}$ nonconstant monomials. The present lifting stores a fixed number of local product variable indices per grid point at a given homotopy order and stencil. These correspond to different truncations, so a resource comparison with a reduced Carleman construction must match the physical observable and its error, as well as the input and output models \citep{liu2021efficient,li2025potential}.

\section{\label{sec:twod}Energy transfer in two-dimensional Taylor--Green flow}

\subsection{Initial pressure balance}

We consider the periodic square $[0,2\pi]^2$ in units of the inverse fundamental wave number $\ell$ and velocity amplitude $U_*$. The initial density is unity and the initial velocity is
\begin{equation}
 \boldsymbol{u}_0=(\sin x\cos y,-\cos x\sin y).
\end{equation}
We set $\gamma=1.4$, $\mathrm{Re}=40$ and $\mathrm{Pe}=25$, where $\mathrm{Pe}=U_*\ell/\chi_0$ uses the internal-energy diffusivity at the reference density. In dimensional variables $\chi_0=\kappa_{\rm dim}/\rho_*$ for the heat-flux law in Eq.~\eqref{eq:closure}. In the units used here, $\rho_*=1$ and $\kappa=\chi_0/(U_*\ell)=1/\mathrm{Pe}$. With the usual isobaric thermal diffusivity $\chi_0/\gamma$, the conventional Prandtl number is $\gamma\mathrm{Pe}/\mathrm{Re}$. The second viscosity is zero. With $\mathrm{Ma}=U_*/c_0$, where $c_0$ is the reference sound speed, the mean pressure in these units is $1/(\gamma\mathrm{Ma}^2)$. The two initialisations have the same velocity, density and mean pressure. The balanced one adds $(\cos 2x+\cos 2y)/4$ to the uniform pressure in the continuum, and the internal energy follows from the perfect-gas closure. At fixed initial density the pressure adjustment also specifies a temperature and entropy perturbation. This cancellation is analogous to data preparation for the periodic barotropic low-Mach limit, where appropriately scaled perturbations make the leading-order acoustic response negligible~\citep{klainermanmajda1981}. Here we impose only the initial longitudinal force balance in a finite-Mach, non-isothermal flow, and subsequent differences reflect the evolution of the complete initial states. In the centred-difference calculations the pressure perturbation is divided by $\cos\Delta x$ so that the pressure gradient exactly balances the second-harmonic advective acceleration generated by the fundamental velocity mode. The ratio of the second-harmonic modified wave number to twice the fundamental one is $\sin(2\Delta x)/(2\sin\Delta x)=\cos\Delta x$, giving a pressure correction of $1+O(\Delta x^2)$.

Let $P_\parallel$ denote the Fourier projection of the velocity onto the wave-vector direction at every nonzero mode, defining the dilatational velocity $\boldsymbol{u}_d=P_\parallel\boldsymbol{u}$. We monitor the spatially averaged specific kinetic energy of the dilatational velocity, $E_\parallel=\tfrac12\langle |P_\parallel\boldsymbol{u}|^2\rangle$ (hereafter the dilatational kinetic energy), which projects $\boldsymbol{u}$ directly, unlike the density-weighted total kinetic energy $K=\tfrac12\langle\rho|\boldsymbol{u}|^2\rangle$. The initial velocity has zero divergence. With uniform pressure the advective term produces the longitudinal acceleration $-(\sin 2x,\sin 2y)/2$ in the continuum, giving $E_\parallel=t^2/8+O(t^3)$ in the present units. The balanced pressure cancels this advective acceleration, in which case $\partial_t\rho|_0=0$ and $\partial_t\boldsymbol{u}|_0=\mu\Delta\boldsymbol{u}_0=-2\mu\boldsymbol{u}_0$ is solenoidal.

The spatially varying balanced pressure at uniform density induces an initial internal-energy gradient through $e_0=p_0/(\gamma-1)$. The energy equation and the perfect-gas law give
\begin{equation}\label{eq:tgv-pressure-rate}
 \left.\partial_t p\right|_0=-\boldsymbol{u}_0\cdot\nabla p_0
 +(\gamma-1)\Phi_0+\kappa\Delta p_0,
 \qquad \Phi_0=4\mu\cos^2x\cos^2y,
\end{equation}
where $\Phi_0=\boldsymbol{\tau}_0:\nabla\boldsymbol{u}_0$ is the initial viscous heating. Differentiating the momentum equation and projecting along the wave vector gives
\begin{equation}\label{eq:tgv-longitudinal-rate}
 \left.P_\parallel\partial_t^2\boldsymbol{u}\right|_0
 =-\nabla\Psi,\qquad \Psi=4\mu p_0+\left.\partial_t p\right|_0.
\end{equation}
The $4\mu p_0$ term comes from viscous decay of both velocity factors in the advective term. Heat conduction changes the initial pressure rate even though the initial dilatation vanishes. In the axial second harmonics $\cos 2x$ and $\cos 2y$, conduction contributes $-\kappa$ to each coefficient of $\Psi$, while velocity decay and viscous heating contribute $\gamma\mu$ in total. At $\mathrm{Re}=40$ and $\mathrm{Pe}=25$ the net coefficient is $-0.005$. Initial pressure advection also generates mixed harmonics at wave numbers $(3,1)$ and $(1,3)$, so longitudinal acceleration at this order persists when $\kappa=0$. The leading longitudinal velocity now grows quadratically, giving $E_\parallel=t^4\langle|\nabla\Psi|^2\rangle/8+O(t^5)$. The continuum coefficient of $t^4$ is $0.0391$ at the benchmark parameters and $0.0397$ when $\kappa=0$ with the same initial fields and viscosity.

\subsection{Reference calculations and finite-order accuracy windows}

The kinetic-energy budget per unit area is
\begin{equation}\label{eq:budget}
 K=\tfrac12\langle\rho |\boldsymbol{u}|^2\rangle,\qquad
 \dot K=\Pi-\mathcal D,\quad
 \Pi=\langle p\nabla\cdot\boldsymbol{u}\rangle,\quad
 \mathcal D=\langle\boldsymbol{\tau}:\nabla\boldsymbol{u}\rangle,
\end{equation}
where brackets denote a spatial average. We monitor the budget defect because the centred stencil does not satisfy a discrete product rule. The reference integrator solves the physical equations directly after dividing the momentum and internal-energy equations by density. The exact two-dimensional lifting applies Eq.~\eqref{eq:shifted-product} along each coordinate direction, and its rowwise residual and final-time comparisons on $8^2$ points use independent time polynomials and include the auxiliary kinetic-energy equation at orders three and four. The physical error curves use $64^2$ centred points, with independent Fourier differentiation and centred grid refinement separating the spatial error from the HAM truncation error. On the centred grid, the maximum absolute defect in the kinetic-energy budget is $2.45\times 10^{-4}$ at $64^2$ over the four cases, whereas the Fourier calculations satisfy the budget to near roundoff on the two finer grids. For the uniform-pressure case at $\mathrm{Ma}=0.4$, the maximum differences in $\Pi$ versus the $128^2$ Fourier result, at seven sample times, are $2.09\times10^{-3}$, $5.45\times10^{-4}$ and $1.38\times10^{-4}$ on $32^2$, $64^2$ and $128^2$ centred grids respectively; the corresponding differences for $\mathcal D$ are $3.46\times10^{-4}$, $8.70\times10^{-5}$ and $2.18\times10^{-5}$. The $64^2$ and $128^2$ Fourier calculations differ in $\Pi$ by $1.02\times10^{-11}$ or less over these samples. Comparing against the same-grid reference thus isolates homotopy truncation even where the remaining spatial error is comparable to the one-percent pressure tolerance. Across both pressure initialisations, the $8^2$ comparison through fourth order  gives a maximum rowwise residual of $4.2\times10^{-17}$ and a final-time vector error of $1.8\times10^{-18}$.

\begin{figure}
\centering
\includegraphics[width=\textwidth]{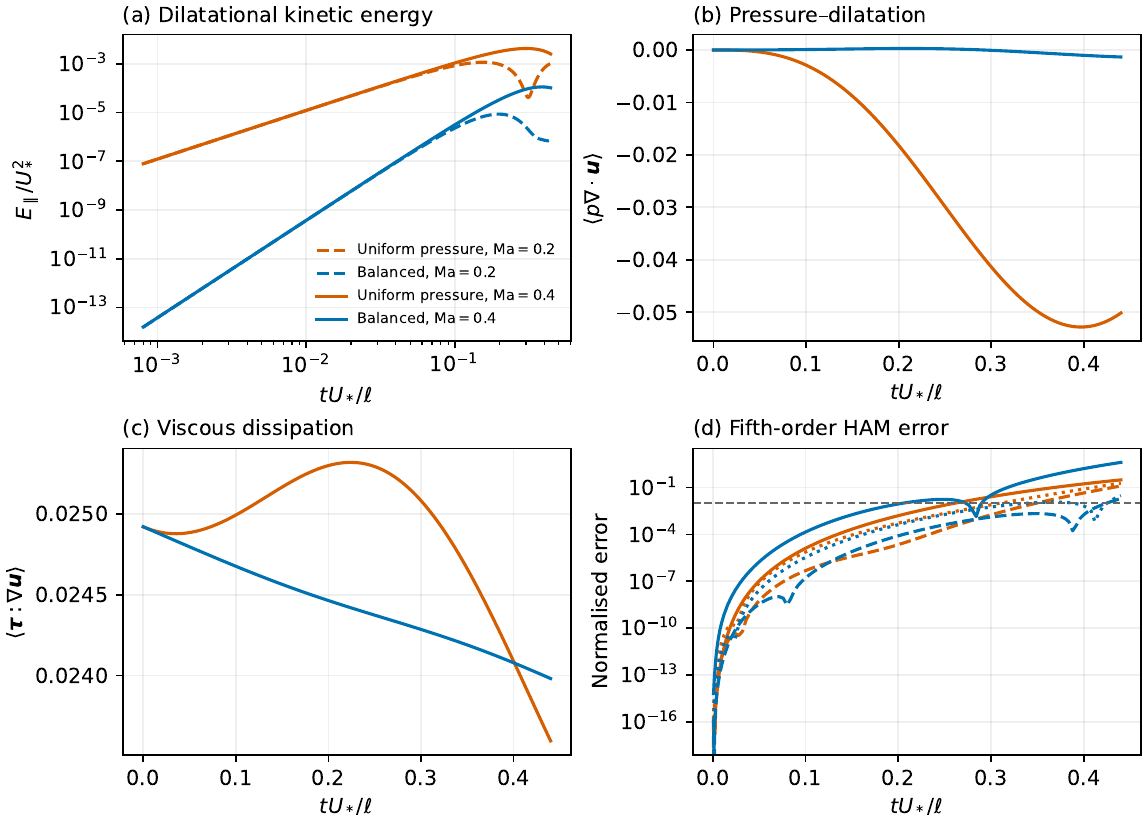}
\caption{\label{fig:tgv}Taylor--Green flow at $\mathrm{Re}=40$, $\mathrm{Pe}=25$. (a) Dilatational kinetic energy $E_\parallel/U_*^2$. Solid and dashed lines denote $\mathrm{Ma}=0.4$ and $0.2$. (b,c) Pressure--dilatation and viscous dissipation at $\mathrm{Ma}=0.4$. (d) Fifth-order HAM errors at $h=-1$ for $\Pi$ (solid), $\mathcal D$ (dashed) and total kinetic energy $K$ (dotted, normalised by the reference decay magnitude), with the one-percent tolerance in grey. Across all panels, orange curves denote uniform initial pressure and blue curves denote balanced initial pressure.}
\end{figure}

Fits over $8\times10^{-4}\le tU_*/\ell\le0.012$ give exponents $1.997$--$1.998$ for uniform pressure and $3.996$--$3.997$ for balanced pressure. \citet{zhang2025taylor} likewise obtained a quadratic onset of density-weighted dilatational kinetic energy for a solenoidal Taylor--Green velocity with a non-uniform, partially balancing initial pressure (second-harmonic coefficient $3/16$ in the present units, versus $0$ for uniform and $1/4$ for balanced pressure), matching the early-time $t^2$ exponent observed here for uniform initial pressure. The balanced initialisation isolates the cancellation of the leading longitudinal acceleration. Figure~\ref{fig:tgv} shows that the smaller dilatational kinetic energy does not imply a longer window of relative accuracy for pressure work. We quantify each accuracy window by the last sampled time before the first crossing of a one-percent tolerance. The error in each observable is divided by its peak absolute reference value over $0\le tU_*/\ell\le0.44$. For $K$ we use the magnitude of the total reference kinetic-energy decay in the denominator, to distinguish an accurate decay curve from an accurate estimate of the dominant initial energy.

\begin{figure}
\centering
\includegraphics[width=\textwidth]{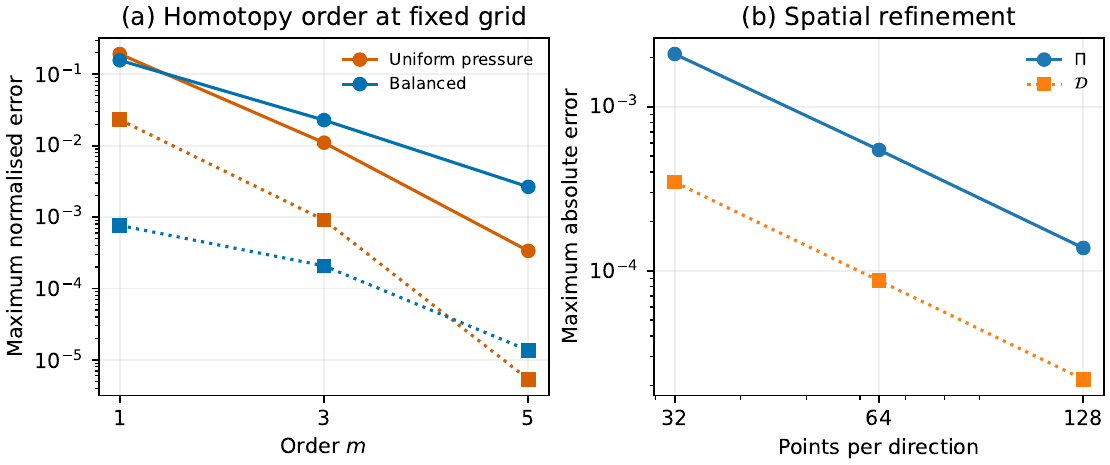}
\caption{\label{fig:tgv-accuracy}Homotopy and spatial error at $\mathrm{Ma}=0.4$. (a) Maximum normalised HAM error relative to the $64^2$ centred reference for $0\le tU_*/\ell\le0.16$ with $h=-1$. Circles show $\Pi$ and squares show $\mathcal D$. (b) Spatial errors against the $128^2$ Fourier reference at seven times for the uniform-pressure case.}
\end{figure}

With $m=5$, $h=-1$, and uniform pressure, the pressure--dilatation windows are $0.144$ and $0.260$ at $\mathrm{Ma}=0.2$ and $0.4$, respectively, while the corresponding viscous-dissipation windows are $0.232$ and $0.348$. With balanced pressure the windows for $\Pi$ are $0.144$ and $0.204$, while those for $\mathcal D$ are $0.280$ and $0.424$. These finite-window results indicate that matching viscous decay can mask loss of the weaker and faster pressure-work signal. Lowering the Mach number at fixed transport coefficients shortens the pressure-work window in the advective units used here. Multiplying these times by $c_0/U_*$ gives acoustic-time windows of $0.72$ and $0.65$ for uniform pressure, and $0.72$ and $0.51$ for balanced pressure, so the difference between the two pressure initialisations at $\mathrm{Ma}=0.4$ persists in acoustic units. A parametric scan in the convergence-control parameter $h$ shows different responses for pressure work and viscous dissipation. At $\mathrm{Ma}=0.4$ and uniform pressure, $h=-0.8,-0.9,-1$ give pressure-work windows of $0.216,0.268,0.260$ and viscous-dissipation windows of $0.372,0.412,0.348$. With balanced pressure, the same values of $h$ give pressure-work windows of $0.204,0.168,0.204$ and viscous-dissipation windows of at least $0.440$, at least $0.440$, and $0.424$; the preferred $h$ therefore depends on the observable and the initial pressure. Figure~\ref{fig:tgv-accuracy} separates order convergence from spatial refinement. For the balanced $\mathrm{Ma}=0.4$ case with $r=1.5$ and  $\delta=0.1$, Theorem~\ref{thm:short} yields explicit fixed-grid convergence bounds, truncated downward to satisfy the theorem inequalities strictly, of $T=2.61\times10^{-3}$, $1.91\times10^{-3}$ and $1.21\times10^{-3}$ at $32^2$, $64^2$ and $128^2$. These bounds guarantee worst-case maximum-norm convergence on the complex ball, whereas the observable windows above measure the time interval over which a given homotopy order remains within the relative error tolerance for the actual smooth flow.

We vary conductivity at fixed $\mathrm{Re}=40$ and $\mathrm{Ma}=0.4$ for the balanced case. The coefficient $\gamma\mu-\kappa$ of each axial second harmonic passes through zero at $\mathrm{Pe}=\mathrm{Re}/\gamma\simeq28.57$. The mixed harmonics keep the continuum coefficient of $t^4$ close to $0.039$ as $\mathrm{Pe}$ increases from $25$ to $80$. On the $64^2$ grid at $m=5$ and $h=-1$, the one-percent pressure-work windows at $\mathrm{Pe}=25$, $28.57$ and $80$ are $0.204$, $0.192$ and $0.180$, while the dissipation windows are $0.424$, $0.424$ and $0.428$. Cancellation of the axial term at onset therefore leaves a substantial separation between the pressure-work and dissipation windows.

Table~\ref{tab:tgv-resources} reports the generator dimension, effective nonzero fraction and unmerged block-encoding bound on the same $64^2$ grid as the physical error curves. At fifth order, identifying identical evolving products reduces the number of auxiliary product blocks by a factor of $5.61$, though including the $4m$ base increments leaves more than $30$ million coordinates. The fraction of nonzero entries decreases with order as the matrix dimension grows, while both the total number of nonzeros and the unmerged stencil bound $\alpha$ increase. Nonzero counts are evaluated after summing coincident stencil-term contributions under the floating-point cancellation tolerance stated in Table~\ref{tab:tgv-resources}, whereas $\alpha$ uses the individual branch bounds of Appendix~\ref{app:circuits}. These figures are classical matrix statistics rather than gate counts; Appendix~\ref{app:program} reports logical gate counts for coarser grids.

\begin{table}
\caption{\label{tab:tgv-resources}Physical Taylor--Green lifting on the $64^2$ centred grid for $h=-1$. The dimension is $N=64^2(4m+C_m)$, where $C_m$ is the compact product-block count. The ranges span the two Mach numbers and both pressure initialisations. Coincident entries are summed before counting nonzeros, with a cancellation tolerance of $64$ machine epsilons times the sum of magnitudes of the coincident terms. The normalisation $\alpha$ uses the separate stencil-term branches before merging.}
\vspace{4pt}
\centering
\small
\begin{tabular}{c r r r c c}
\toprule
 & \multicolumn{2}{c}{Product blocks} & & \multicolumn{2}{c}{Compact generator ranges}\\
\cmidrule(lr){2-3}\cmidrule(lr){5-6}
$m$ & Unreduced & Compact & Compact dimension $N$ & $\mathrm{nnz}(A_m)/N^2$ & Unmerged $\alpha$\\
\midrule
3 & 40 & 26 & $155\,648$ & $[3.49,3.58]\times10^{-5}$ & $[4.94,10.49]\times10^3$\\
4 & 1408 & 518 & $2\,187\,264$ & $[1.82,1.85]\times10^{-6}$ & $[5.44,12.91]\times10^5$\\
5 & $42\,064$ & 7500 & $30\,801\,920$ & $[1.63,1.66]\times10^{-7}$ & $[2.62,6.40]\times10^7$\\
\bottomrule
\end{tabular}
\end{table}

\begin{figure}
\centering
\includegraphics[width=0.9\textwidth]{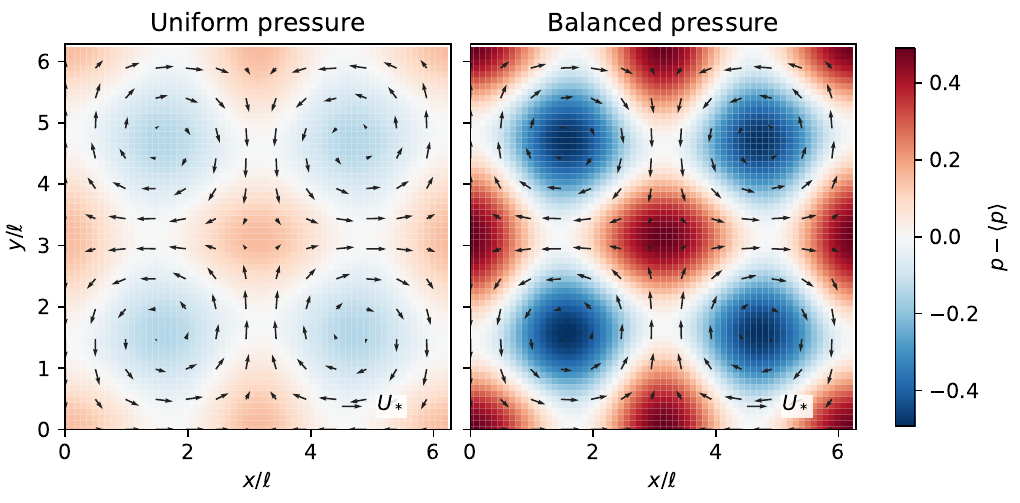}
\caption{\label{fig:flow}Taylor--Green flow at $tU_*/\ell=0.16$ and $\mathrm{Ma}=0.4$. Colour shows pressure relative to the spatial mean. Arrows show the velocity at every fifth grid point, with reference speed $U_*$. Both panels use the same scales.}
\end{figure}

\section[Finite-polynomial final-time solution and its quantum implementation]{\label{sec:cost}Finite-polynomial final-time solution and its quantum implementation}

\begin{theorem}[Finite-polynomial solution at the final time]\label{thm:poly}
For product lifting with $L=\partial_t$, the fixed-grid affine equation is $\dot y=A_m y+b_m$, whose generator matrix $A_m$ is strictly block lower-triangular when ordered by homotopy order $1,\dots,m$ and therefore nilpotent with $A_m^m=0$. For every $T\ge0$, its exact solution at time $T$ is given without time-stepping by
\begin{equation}\label{eq:terminal}
 y(T)=\sum_{j=0}^{m-1}\frac{T^j}{j!}A_m^j y(0)
       +\sum_{j=0}^{m-1}\frac{T^{j+1}}{(j+1)!}A_m^j b_m.
\end{equation}
\end{theorem}

\begin{proof}
Every nonzero matrix entry of $A_m$ couples a column of strictly lower positive homotopy order to a row of larger order, and the highest order is $m$. Hence no nonzero product of $m$ entries exists, so $A_m^m=0$. Substitution of the finite matrix exponential $e^{TA_m}=\sum_{j=0}^{m-1}T^j A_m^j/j!$ into the variation-of-constants formula and integration of each monomial gives Eq.~\eqref{eq:terminal}.
\end{proof}

Suppose $A_m$ has a block encoding with normalisation $\alpha$ and that the normalised initial and source vectors can be prepared, with their norms known. Products of $j$ encodings have normalisation $\alpha^j$. The two sums in Eq.~\eqref{eq:terminal} can be applied by linear combinations of unitaries (LCU)~\citep{berry2015taylor}, with a state-preparation branch for each nonzero input and a \texttt{SELECT} branch for each retained power. Projecting onto the subspace with all success ancillas in $|0\rangle$ yields the subnormalised vector $y(T)/\mathcal Z$, where

\begin{align}\label{eq:normalisation}
\mathcal Z={}&\|y(0)\|_2\sum_{j=0}^{m-1}\frac{(\alpha T)^j}{j!}\nonumber\\
 &+T\|b_m\|_2\sum_{j=0}^{m-1}\frac{(\alpha T)^j}{(j+1)!} .
\end{align}

In the exact-circuit model, post-selection onto the all-zero ancilla outcome succeeds with probability $p=\|y(T)\|_2^2/\mathcal Z^2$. The dimensionless product $x=\alpha T$ governs the weights of powers in Eq.~\eqref{eq:normalisation}. For fixed $m\ge2$ and $x\gg m$, the highest-degree term gives
\begin{equation}\label{eq:large-alpha-time}
 \mathcal Z\sim x^{m-1}
 \left(\frac{\|y(0)\|_2}{(m-1)!}+\frac{T\|b_m\|_2}{m!}\right).
\end{equation}
For example, the fifth-order compact one-dimensional encoding in Table~\ref{tab:compact} gives $x\simeq 1.78\times10^3$ and $p\simeq 1.45\times10^{-21}$ at $T=0.02$ after inserting the source and final-time norms. At fixed ratio of the final-time norm to the factor in parentheses, the preparation probability falls as $x^{-2(m-1)}$, while changes in the final-time norm must be accounted for when that ratio varies. Independent post-selection takes $1/p$ calls per accepted state on average, and with reversible input preparation, amplitude amplification takes $O(1/\sqrt{p})$ calls for a nonzero target. Each call in the explicit serial \texttt{SELECT} construction uses $O(m^2)$ controlled encoding invocations.

In the fifth-order $64^2$ Taylor--Green lifting at $T=0.20$, the unmerged stencil-term sum gives $\alpha T$ between $5.24\times10^6$ and $1.28\times10^7$. The finite polynomials give $\|b_m\|_2=28.78$--$32.12$ and $\|y(T)\|_2=45.40$--$86.50$, with $y(0)=0$, and the resulting exact-circuit success probabilities are given in Table~\ref{tab:tg-normalisation}. Choosing $\alpha\tau\le1$ would require $5.24$--$12.80$ million segments with this encoding. Exact affine evolution composes over those segments without time-stepping discretisation error, though the overall cost must include intermediate state preparation or coherent reuse together with their success probabilities.

\subsection{Reducing the normalisation}

Stencil terms with identical row index, column index and periodic shift contribute to the same matrix entries. Combining their grid coefficients $f_g(p)$ first gives a single weighted-shift branch per group $g$, with normalisation $\alpha_{\rm merged}=\sum_g\max_p|f_g(p)|$. The topology of the original stencil terms can still be used, while the value oracle now evaluates the merged sum or accesses its table. Given coherent row-location, column-location and finite-precision value oracles together with a known entry bound $M\ge\max_{i,j}|(A_m)_{ij}|$, a sparse-access block encoding provides the alternative normalisation $\alpha_{\rm sparse}=\sqrt{s_r s_c}\,M$~\citep{gilyen2019qsvt}. Here $s_r$ and $s_c$ bound the maximum number of nonzeros per row and per column  by counting the merged shifts incident to each variable index, and the cost of turning these lists and values into coherent row and column access must be charged.

The strict block lower-triangular structure of $A_m$ also restricts the matrix products needed in Eq.~\eqref{eq:terminal}. While each deformation equation depends on lower orders, acting with $A_m$ on a vector moves weight along paths of strictly increasing homotopy order ($s\to r$ with $r>s$). Let $B_{rs}$ sum the merged branch bounds coupling source order $s$ to target order $r>s$, and let $\beta_s$ be the norm of the order-$s$ component of $b_m$. Block-encoding each pair of orders separately and taking an LCU over paths of strictly increasing homotopy order gives, for $y(0)=0$,
\begin{equation}\label{eq:rank-lcu}
 \mathcal Z_{\rm rank}=T\|b_m\|_2+
 \sum_{j=1}^{m-1}\frac{T^{j+1}}{(j+1)!}\,\boldsymbol{1}^{\mathsf T}B^j\boldsymbol{\beta}.
\end{equation}
The zeroth power uses the full source, and the other terms prepare the corresponding order-restricted source components. Each entry of $B^j$ bounds the sum of encoding weights along paths of length $j$, which avoids applying a large bound from a high-order block at every power. This requires \texttt{SELECT} over order paths and the ability to prepare order-restricted sources. The coefficient and probability estimates in Table~\ref{tab:tg-normalisation} assume exact oracle implementations for these operations.

\begin{table}
\caption{\label{tab:tg-normalisation}Generator block-encoding factor $\alpha$, polynomial LCU scale and single-interval all-zero-ancilla success probability at $m=5$ and $T=0.20$ on the $64^2$ Taylor--Green grid. Ranges span the four combinations of Mach number and initial pressure. Probabilities use the actual final-time and source norms. For $y(0)=0$, the merged order-path construction normalises the source-integral polynomial in Eq.~\eqref{eq:terminal} directly via Eq.~\eqref{eq:rank-lcu}. All values are computed from the stencil terms and independent time polynomials.}
\vspace{4pt}
\centering
\small
\setlength{\tabcolsep}{5pt}
\begin{tabular}{lccc}
\toprule
Block-encoding scheme & Generator factor $\alpha$ & Polynomial LCU scale & Exact-circuit success probability\\
\midrule
Factorised stencil terms & $[2.62,6.40]\times 10^7$ & Eq.~\eqref{eq:normalisation} ($\mathcal Z$) & $[1.25\times10^{-51},\,1.35\times10^{-48}]$\\
Merged coefficients & $[3.39,6.97]\times 10^5$ & Eq.~\eqref{eq:normalisation} ($\mathcal Z$) & $[6.29\times10^{-36},\,1.71\times10^{-33}]$\\
Sparse access & $[2.15,9.26]\times 10^5$ & Eq.~\eqref{eq:normalisation} ($\mathcal Z$) & $[6.51\times10^{-37},\,6.48\times10^{-32}]$\\
Merged order paths & Not used separately & Eq.~\eqref{eq:rank-lcu} ($\mathcal Z_{\rm rank}$) & $[1.10\times10^{-25},\,2.75\times10^{-23}]$\\
\bottomrule
\end{tabular}
\end{table}

Merging lowers $\alpha$ by up to about two orders of magnitude in these cases, and restricting the products to  paths of strictly increasing homotopy order yields a further improvement of several orders of magnitude in success probability. Even that exact-circuit probability calls for over $10^{11}$ coherent calls at the best of the tested fifth-order points if the objective is preparing the full final-time quantum state by amplitude amplification.

\subsection{Circuit implementations and observable cost}

The factorised stencil implementation encodes $A_m$ without assembling its matrix and obtains its powers by repeated calls, whereas directly encoding a precomputed polynomial combines cancellations and coincident contributions before normalisation. For $y(0)=0$ at third order, Appendix~\ref{app:program} groups the entries of $P=I+TA_3/2+T^2 A_3^2/6$ by bitwise XOR mask $i\oplus j$ (each group supported on a permutation matrix) to form a direct block encoding. Its normalisation is $1.462$ and its all-zero-ancilla success probability is $0.469$ for the tested source. On the same grid at the same time, the factorised stencil implementation gives $\alpha_A=90.65$ for $A_3$ and an effective polynomial normalisation $\sum_{j=0}^2(\alpha_A T)^j/(j+1)! = 2.454$, corresponding to an exact-circuit success probability of $0.166$. At larger orders the powers can couple more variable pairs and accumulate longer stencil shifts, so direct polynomial encoding must then  account for constructing, storing and loading the polynomial coefficients. By contrast, the factorised implementation charges the stencil-term queries and  repeated calls while inheriting the product normalisations in Eq.~\eqref{eq:normalisation}. At the same third-order grid, final time and signed observable, the tabulated-polynomial and factorised-stencil circuits use $25\,866$  and $294\,491$ logical Toffoli gates, respectively. Appendix~\ref{app:program} compares these implementations and gives a same-observable comparison with midpoint-quadrature LCHS.

The stencil-term branches obtain their coefficients from zeroth-order fields and local grid functions derived from them. In \texttt{oracq}~\citep{oracq2026paper}, a modular, typed Python framework for quantum scientific computing, separately bindable oracle interfaces for block-encoding the generator and preparing input states decouple the algorithm from the underlying data-access primitive. Concrete implementations can therefore use explicit \texttt{QROM}-style tables of address-controlled rotations, general sparse-matrix access via compatible hardware \texttt{QRAM}~\citep{chen2023sparseqls}, or reversible arithmetic for analytic profiles  evaluated from the grid address and a finite stencil at the chosen precision. Each implementation must account for data construction and loading, state preparation, reversible evaluation and gate synthesis. Appendix~\ref{app:circuits} describes the gate-based stencil implementation used for the logical counts.

Physical reconstruction sums the base increments through a linear map and adds the known initial field. Unlike reduced-order quantum models targeting flow-field reconstruction and classification from existing data~\citep{fang2025polyqrom}, we directly estimate a single signed velocity Fourier mode at the final time. Estimating a bounded-variance observable on the normalised state to additive precision $\eta$ at constant confidence takes $O(\eta^{-2})$ post-selected copies by sampling, and recovering physical values also requires the state norm and the norm of the linear functional. For the signed linear readout, repeated Hadamard tests for a physical tolerance $\epsilon$ cost $O(G_{\rm shot}\mathcal Z^2\|Q^{\mathsf T}\ell\|_2^2/\epsilon^2)$ logical gates at fixed confidence, where $G_{\rm shot}$ includes preparation, \texttt{SELECT} and the probe circuit. A classical sparse implementation of the same affine polynomial uses at most $m$ matrix--vector products. Any advantage for this observable requires the total quantum cost, including oracle construction and the accuracy of the homotopy approximation at the requested resolution, to be smaller than the cost of evaluating the observable classically.

For any nonzero nilpotent matrix $A_m$, the Hermitian part is traceless and nonzero, so it has a positive eigenvalue. In the original coordinate norm the homogeneous propagator $e^{TA_m}$ thus cannot be passed directly to the dissipative LCHS identity~\citep{an2023lchs,an2026lchs}. For the comparison in Appendix~\ref{app:program}, we append a constant coordinate equal to one, place $b_m$ in the last column of the enlarged generator, and shift the enlarged matrix by the largest eigenvalue of its Hermitian part, which introduces an exponential rescaling of the final amplitude. Evaluating Eq.~\eqref{eq:terminal} directly uses the original stencil terms and source-state preparation while retaining the nilpotence of $A_m$.

\section[Scope and conditional all-time convergence]{\label{sec:boundary}Scope and conditional all-time convergence}

The smooth periodic benchmarks isolate time homotopy truncation by comparison with same-grid solutions of the non-isothermal equations. Periodicity gives periodic finite-difference stencils and a Fourier basis for longitudinal projection and for diagnosing the null space of the grid derivative. The Taylor--Green cases vary the initial pressure balance and Mach number while holding the velocity and transport coefficients fixed. The short-time estimate applies on a ball with positive density; strong gradients can enlarge its fixed-grid Lipschitz bound and shorten the explicit convergence interval. Kinetic-energy-consistent split discretisations of the convective fluxes~\citep{pirozzoli2010split} provide a route to reducing the budget defect. Equation~\eqref{eq:shifted-product} applies to any finite shift stencil, including polynomial split fluxes after expanding their local products, though the resulting semidiscrete hierarchy would have to be derived for those fluxes. Changing the divergence and gradient pairing, for instance by staggering, can remove the alternating (checkerboard) pressure-gradient null mode, whereas replacing only the viscous and thermal second derivatives leaves this first-derivative null mode in the density--velocity coupling. Any replacement stencil requires its own consistency and stability analysis.

The homotopy-order grading covers all derivative-bearing polynomial residuals for the compressible perfect-gas system and its Euler limit. For example, time-independent, local affine ghost-cell rules can be substituted at the homotopy order of the relevant increment, with prescribed data in order zero as in Sec.~\ref{sec:defo}. This preserves the strict lower-order structure and hence the nilpotence of $A_m$ whenever the resulting boundary stencils retain the lower-order dependence of the interior equations, though boundary stencil terms and their access costs must then be recomputed. A pressure projection, a tabulated constitutive law and vacuum require separate analyses.

A dissipative auxiliary operator offers a route toward a time-independent sufficient condition. Appendix~\ref{app:convergence} formulates this route in terms of an inverse bound in the same trajectory spaces used to control the nonlinear remainder. An acoustic energy with diagonal density, velocity and internal-energy weights does not immediately give a strict negative logarithmic norm, because a mean-zero pure density perturbation has zero instantaneous dissipation in this norm. Compensated energies can transfer viscous and thermal dissipation to density through the acoustic coupling~\citep{shizuta1985}. On the even periodic grid here, the alternating mode $(-1)^p$ also lies in the kernel of both $D$ and the composite second-derivative operator $D^2=D\circ D$; it survives removal of the constant mode and prevents a strictly decaying semigroup on the full mean-zero subspace. A uniform global estimate for a revised scheme would therefore have to address the exact discrete spectrum and the nonlinear treatment of conserved mean modes. Changing the auxiliary operator introduces same-order coupling terms, so the closure and final-time evaluation need their own analysis for that extension.

\section{\label{sec:disc}Conclusions}

Product lifting preserves the finite HAM dynamics of the non-isothermal compressible Navier--Stokes equations on the invariant set where every auxiliary variable equals its defining product. Tracking higher-order generated time derivatives and applying Eq.~\eqref{eq:shifted-product} makes this construction verifiable component by component against the semidiscrete hierarchy, while Theorem~\ref{thm:short} provides an explicit fixed-grid convergence bound linking the finite hierarchy to the physical semidiscrete solution. Because the lifted generator matrix $A_m$ is strictly block lower-triangular in homotopy order, it is nilpotent ($A_m^m=0$) and yields the finite-polynomial final-time solution~\eqref{eq:terminal} together with a modular quantum implementation verifiable amplitude by amplitude. In the two-dimensional Taylor--Green benchmarks, the kinetic-energy budgets clearly separate the relative accuracy window for pressure work from that for viscous dissipation: initial pressure balance delays the onset of the dilatational kinetic energy $E_\parallel$ from quadratic to quartic in time, thermal diffusion modifies the axial harmonics of the delayed longitudinal acceleration, and pressure advection excites persistent mixed harmonics, so an accuracy check based solely on kinetic-energy decay would miss the shorter pressure-work window. For the finite-order semidiscrete flow, evaluating Eq.~\eqref{eq:terminal} directly recovers a signed velocity Fourier mode at the final time without introducing time-stepping discretisation error. Merging coincident stencil coefficients and combining block-encoded products along paths of strictly increasing homotopy order substantially lower the normalisation cost of this lifting, though fifth-order evolution on the fine grid still incurs a substantial state-preparation cost; establishing grid-uniform control of the homotopy error and reducing the costs of block encoding, data preparation and observable readout remain the next steps towards practical quantum simulation of compressible flows.

\appendix

\renewcommand{\theHsection}{app.\Alph{section}}

\section{\label{app:circuits}From stencil descriptors to a block encoding}

In the circuit implementation, each product variable index of Sec.~\ref{sec:lift} is stored as a label (called a signature in the implementation), and each stencil term is represented by a five-tuple stencil descriptor.

Use the tensor index $(\nu,p)$ for a signature $\nu$ and a grid point $p$. Each descriptor gives a row signature $\nu$, a column signature $\nu'$, an offset $\eta$, a scalar $a$, and a list of known factors $g_i$. Its location and value contributions are
\begin{equation}\label{eq:oracle-edges}
 \ell_{\rm col}=(\nu',p+\eta\pmod n),\qquad
 v=a\prod_i g_i(p).
\end{equation}
In several dimensions the periodic shift is applied componentwise, and the list of factors includes their own offsets and derivative orders. A row query enumerates the descriptors of $\nu$, sums contributions to the same matrix entry, and skips terms that cancel to zero. A column query reverses the shift and interchanges row and column index roles. Sources are evaluated from the terms with no evolving factor, including the first-order time derivatives. Equation~\eqref{eq:convection-entry} has $\eta=0$ and a single density factor, whereas the two entries for $\rho^{(0)}u^{(0)}D u^{(2)}$ use $\eta=\pm1$ and place the signed difference weights in $a$. Substituting a time-differentiated linear coupling recursively composes the corresponding descriptors, adding the shifts and moving the evaluation points of its data factors.

The compiled quantum circuit uses these descriptors directly in the block encoding. For each descriptor, a reversible branch tests the input index $\nu'$, performs the inverse grid shift, and maps the index to $\nu$. Controlled single-qubit rotations encode the amplitudes $g_i(p)/B_i$, with $B_i=\max_p|g_i(p)|$, onto the $|0\rangle$ state of dedicated success ancilla qubits whose joint $|0\rangle$ outcome gives the product amplitude; branches with a zero factor are dropped. Applying LCU over descriptors, including the sign of $a$, gives a block encoding with $\alpha=\sum_{\rm edges}|a|\prod_i B_i$, with unused indices and register entries padded to the next power of two flagged. At fixed order and stencil pattern, the descriptor list before pruning zero factors is independent of the number of grid points. Each data factor is a grid function with one value per point, such as a zeroth-order field, a spatial derivative or an absorbed first-order time derivative computed from the zeroth-order data. The factorised stencil implementation evaluates the branches without assembling or storing the lifted matrix. The order-two circuits in Appendix~\ref{app:program} use point-controlled rotations as explicit gate tables for the factors and gate-based preparation of the source, and their logical counts include these gates. A compatible hardware \texttt{QRAM} or reversible analytic evaluator can implement the same branch amplitudes through the modular interface, with its own data, precision and gate costs.

\section{\label{app:program}Circuit execution and LCHS comparison}

Within the modular \texttt{oracq} framework introduced in Sec.~\ref{sec:cost}~\citep{oracq2026paper}, Eq.~\eqref{eq:terminal} is represented by a program with declared oracle interfaces (called slots in \texttt{oracq}) for the generator block encoding and for preparing each nonzero input state. We instantiate these interfaces with the concrete descriptor-generated block encoding and gate-based preparation of the evaluated source vector. For the two order-two examples, $y(0)=0$, so only the source state-preparation interface is required. This order tests a nonzero $A_m b_m$, the two-term source-integral polynomial, and the reversible encoding of signed coefficients, while the order-three and higher closure tests in Table~\ref{tab:numeric} test the auxiliary product variables. The reference simulator checks the rescaled  amplitudes on the all-zero ancilla subspace against independent time polynomials on the full register padded to a power of two. The resource estimates in Table~\ref{tab:circuits} are logical gate counts for these concrete circuit implementations.

\begin{table}
\caption{\label{tab:circuits}Concrete logical circuit simulation at $m=2,T=0.02$ for the physical equations. Here $p$ is the all-zero-ancilla success probability, and Error is the maximum absolute deviation of rescaled final-time amplitudes.}
\vspace{4pt}
\centering
\small
\begin{tabular}{rrrrrrr}
\toprule
$n$ & Dimension & Qubits & Toffoli & $\alpha$ & $p$ & Error\\
\midrule
4 & 24 & 14 & 4158 & 14.44 & 0.764 & $1.73\times 10^{-18}$\\
8 & 48 & 15 & 7620 & 28.75 & 0.604 & $3.65\times 10^{-18}$\\
\bottomrule
\end{tabular}
\end{table}

\subsection{Third-order circuit and a signed physical observable}

The compact lifting at $m=3$ and $n=4$ uses nine base-increment blocks and the ten compact product blocks reported in Table~\ref{tab:compact}. Each block holds four grid values, giving $(9+10)\times4=76$ values, and in this test $y(0)=0$. We assemble the sparse coefficients of the matrix polynomial $P=I+T A_3/2+T^2 A_3^2/6$ at $T=0.02$ and group its entries by bitwise XOR mask $i\oplus j$ (each group supported on a permutation matrix) to construct a direct block encoding. This tabulated coarse-grid implementation incurs classical preprocessing and sparse storage. We also compile the factorised stencil implementation at the same order, grid and final time. The third-order circuit takes the prepared $b_3$, so the desired final-time vector is $T P b_3$; it tests propagation involving the actual lifted products without preparing the final solution as input.

Let $Q$ be the linear map that sums the velocity-increment blocks, and let $\ell_p=(2/n)\sin(2\pi p/n)$. The physical observable is $\ell^{\mathsf T}(u_0+Qy(T))$. In the circuit, a separate probe state is prepared proportional to $Q^{\mathsf T}\ell$, and its inverse preparation maps the selected inner product onto the amplitude of the all-zero computational basis state. A Hadamard  test with a controlled circuit converts this real signed inner product to the difference of two ancilla probabilities. If $\mathcal Z$ denotes the chosen circuit normalisation, multiplication by $\mathcal Z\|Q^{\mathsf T}\ell\|_2$ gives the signed increment, to which we add the known initial coefficient. Unlike direct basis sampling, this procedure retains the sign. Sampling the Hadamard test ancilla  to standard error $\epsilon$ requires a number of runs proportional to $\epsilon^{-2}$, with the physical error carrying the same normalisation multiplier.

The simulated coefficient from the tabulated polynomial circuit is $0.1005431827$, agreeing with the independent time polynomial to the displayed precision. The matrix entries from the factorised descriptors agree with the assembled rows to $2.78\times10^{-16}$. Projecting the  simulated descriptor branches onto the all-zero ancilla subspace and summing the rescaled contributions agrees with the independent final-time vector to $1.08\times10^{-19}$. Table~\ref{tab:matched-quantum} gives resource counts including the source and probe. The factorised circuit uses three controlled generator calls in one serial \texttt{SELECT}, one for the linear branch and two for the quadratic branch. The tabulated circuit applies one encoding of $P$ and stores 546 nonzero polynomial coefficients in preprocessing.

\begin{table}
\caption{\label{tab:matched-quantum}Logical resources per signed readout at $m=3$, $n=4$ and $T=0.02$. Work ancillas count the additional decomposition ancillas reported for multiply controlled gates. Rotations are listed before synthesis to a discrete gate set. The rescaling multiplies the Hadamard test expectation value to give the increment. The probability for the factorised implementation uses the independently evaluated final-time norm and the circuit normalisation.}
\vspace{4pt}
\centering
\small
\begin{tabular}{lrrrrcc}
\toprule
Circuit implementation & Qubits & Work ancillas & Toffoli & Rotations & $p$ & Rescaling\\
\midrule
Tabulated polynomial & 16 & 14 & $25\,866$ & 960 & $0.469$ & $0.02512$\\
Factorised descriptors & 36 & 16 & $294\,491$ & $6290$ & $0.166$ & $0.04218$\\
\bottomrule
\end{tabular}
\end{table}

\subsection{A two-dimensional readout}

On an $8^2$ Taylor--Green grid with balanced initial pressure at $\mathrm{Ma}=0.4$, we compile the order-two physical lifting using separate periodic shifts for the two coordinates. The probe extracts the signed coefficient of $\sin(3x+y)$ in the longitudinal velocity along the centred-difference wave vector, which points along $(1,1)$ on this grid. At $T=0.02$ the independent deformation polynomial gives $-2.8658\times10^{-5}$, consistent with the delayed mixed-mode response derived in Sec.~\ref{sec:twod}. The factorised circuit has 512 lifted coordinates and uses 21 logical qubits, 14 additional gate-decomposition ancillas, $144\,238$ Toffoli gates and 4074 unsynthesised rotations per signed readout. The all-zero-ancilla success probability inferred from the final-time norm is $0.3654$. The descriptor matrix agrees with the independent stencil rows to $4.44\times10^{-16}$. Simulating each implemented descriptor branch and projecting onto the all-zero ancilla subspace gives the same signed coefficient and a maximum final-time vector error of $2.10\times10^{-19}$. The gate count is for the complete Hadamard circuit, whereas the numerical check uses branch projection and contraction.

\subsection{Comparison with midpoint-quadrature LCHS}

For LCHS we use the same $m=3,n=4$ lifting, final time $T=0.02$ and signed velocity coefficient. Appending a constant coordinate equal to one (placing $b_m$ in the last column of the enlarged generator) yields a largest Hermitian-part eigenvalue $\sigma=3.6573$ and requires undoing the shift by $e^{\sigma T}$. We integrate the Cauchy-kernel identity~\citep{an2023lchs} using the midpoint rule after the substitution $k=\tan(\pi x)$, evaluating each Hamiltonian branch by classical diagonalisation. The absolute signed-coefficient errors at $2048$, $8192$ and $16\,384$ quadrature nodes are $1.32\times10^{-4}$, $1.01\times10^{-5}$ and $2.81\times10^{-6}$, where the last is the first tested quadrature node count below the target quadrature error $10^{-5}$ (at fixed spatial grid $n=4$). That quadrature would use $16\,384$ controlled Hamiltonian simulations per serial \texttt{SELECT}, with maximum $T\|H_k\|_2\simeq1.53\times10^3$. A controlled parameterised simulator could multiplex these branches, and its cost would depend on the Hamiltonian oracle, the range of $k$ and the synthesis accuracy. The observable Hadamard rescaling is $1.318$ for this shifted construction. These are branch counts and scales for the specified quadrature; improved LCHS kernels~\citep{an2026lchs}, alternative Hamiltonian simulation or a different source representation can change them, and a gate-count comparison would require concrete circuit implementations for those operations. FOQHAM and Carleman lifting would additionally require a shared physical discretisation, matched observable error and explicit input oracles.

\section{\label{app:convergence}Convergence proof and a conditional dissipative extension}

\begin{proof}[Proof of Theorem~\ref{thm:short}]
With $U^{(0)}$ stationary, the zeroth-order deformation reads
\begin{equation}
 D(q,U)\dot U=-qh F(U),\qquad D(q,U)=I-q(I+hM(U)).
\end{equation}
For $|q|\le r$ and $\|U-U^{(0)}\|_\infty\le\delta$, one has $\|q(I+hM(U))\|_\infty\le r(a_0+|h|\delta)$. The Neumann series gives $\|D^{-1}\|_\infty\le d_0^{-1}$. The resolvent identity and $\|M(U)-M(V)\|_\infty\le\|U-V\|_\infty$ give $\|D(q,U)^{-1}-D(q,V)^{-1}\|_\infty\le r|h|d_0^{-2}\|U-V\|_\infty$. Therefore the deformed vector field has norm at most $B$ and Lipschitz constant at most $K$ on the ball.

Picard iteration of the time integral from $U^{(0)}$ maps the trajectory ball into itself since $TB\le\delta$, and contracts since $TK<1$. The estimates are uniform on the complex $q$ disk. The iterates are holomorphic in its interior and continuous up to the boundary, as is the uniform limit. At $q=0$ the limit is the time-independent $U^{(0)}$. Matching Taylor coefficients yields Eq.~\eqref{eq:defo} with homogeneous initial increments. Cauchy estimates on the time-supremum norm, using radii approaching $r$, give $\sup_t\|U^{(j)}(t)\|_\infty\le\delta r^{-j}$ for $j\ge1$. At $q=1$ the equation reduces to $M(U)\dot U=F(U)$. The solution is real by uniqueness and remains in the positive-density ball. Summing the Cauchy bounds gives Eq.~\eqref{eq:short-tail}.
\end{proof}

\begin{proposition}[Uniform dissipative criterion]\label{prop:diss}
Let $w=U-U^{(0)}$ have homogeneous initial data, and split the residual as $N(U^{(0)}+w)=N(U^{(0)})+Lw+\mathcal R(w)$ where $\mathcal R(0)=0$ and $L=\partial_t-\mathcal A$. On a complex trajectory ball, assume that $L^{-1}\mathcal R$ is holomorphic and Lipschitz with constant $K_*$, and that $\|L^{-1}N(U^{(0)})\|\le D_*$. Assume these bounds hold uniformly in the window length. If some $r>1$ and $\delta>0$ satisfy
\begin{equation}\label{eq:diss-test}
 r R_*<1,\qquad r|h|D_*+rR_*\delta\le\delta,
 \qquad R_*=|1+h|+|h|K_* ,
\end{equation}
then the modified HAM series has the geometrically bounded tail of Eq.~\eqref{eq:short-tail} in the chosen trajectory norm on each finite window, with constants independent of the window length.
\end{proposition}

\begin{proof}
Applying $L^{-1}$ to the zeroth-order deformation gives
\begin{equation}
 w=q\left[(1+h)w+hL^{-1}N(U^{(0)})+hL^{-1}\mathcal R(w)\right].
\end{equation}
For $|q|\le r$, Eq.~\eqref{eq:diss-test} gives self-mapping and contraction on the trajectory ball. Uniform Picard convergence gives holomorphic dependence on $q$. The Cauchy argument of Theorem~\ref{thm:short} then applies with the assumed trajectory norm.
\end{proof}

For example, on a stable subspace on which $\|e^{t\mathcal A}\|\le C_s e^{-\sigma t}$ in a common state norm, the Duhamel operator on continuous forcings has norm at most $C_s/\sigma$. The bound follows by integrating the exponential kernel, and contains no factor of $T$. If the nonlinear remainder maps this trajectory space into the forcing space with Lipschitz constant $L_{\mathrm{rem}}$, then $K_*\le C_s L_{\mathrm{rem}}/\sigma$. For a quasilinear continuum problem, an appropriate smoothing or maximal-regularity estimate must justify this mapping in the presence of derivative terms. A spectral gap alone does not supply that estimate.

To see where the dimensionless flow numbers enter, write $\mathrm{Re}=U_*\ell/\nu$, $\mathrm{Pe}=U_*\ell/\chi$, and $\mathrm{Ma}=U_*/c_s$ for velocity scale $U_*$, length $\ell$, kinematic viscosity $\nu$, thermal diffusivity $\chi$, and reference sound speed $c_s$. If a particular discretisation and trajectory ball admit $\sigma\ge c_d\min(\nu,\chi)/\ell^2$ and $L_{\mathrm{rem}}\le C_N U_*/\ell$, then $K_*\le (C_s C_N/c_d)\max(\mathrm{Re},\mathrm{Pe})$. The constants may depend on Mach number, perturbation regularity, density bounds, and the discrete energy. This is the sense in which a dissipation-to-nonlinearity ratio analogous to the Carleman criterion~\citep{liu2021efficient} would arise. Those two estimates and the self-mapping bound must be proved for the chosen scheme to assert a particular dimensionless threshold. The even-grid obstruction in Sec.~\ref{sec:boundary} rules out the stated positive gap on its full mean-zero subspace.

\bibliographystyle{unsrtnat}

\bibliography{refs}

\end{document}